\documentclass[conference]{IEEEtran}
\usepackage{fancyhdr}
\usepackage{amsmath,amssymb}
\usepackage{amsfonts}
\usepackage{xcolor}
\usepackage[dvips]{graphicx}
\usepackage{dsfont}
\usepackage{balance}
\usepackage{algpseudocode}
\usepackage{algorithm}
\usepackage{enumerate}
\usepackage{graphics}
\usepackage{subfigure}
\usepackage{cite}

\newtheorem{theorem}{Theorem}

\newtheorem{proposition}{Proposition}

\newcommand{\argmax}{\arg\!\max}
\begin{document}

\title{Simultaneous Lightwave Information and Power Transfer (SLIPT) for Indoor IoT Applications}
\author{\IEEEauthorblockN{
Panagiotis D. Diamantoulakis\IEEEauthorrefmark{1} and George K. Karagiannidis\IEEEauthorrefmark{1}}

\IEEEauthorblockA{\IEEEauthorrefmark{1}Department of Electrical
and Computer Engineering, Aristotle University of Thessaloniki,
GR-54124 Thessaloniki,
        Greece}

\IEEEauthorblockA{e-mails: \{padiaman, geokarag\}@auth.gr}
}
\maketitle

\begin{abstract}
We present the concept of Simultaneous Lightwave Information and Power Transfer (SLIPT) for indoor Internet-of-Things (IoT) applications. Specifically,  we propose novel and fundamental SLIPT strategies, which can be implemented through Visible Light or Infrared communication systems, equipped with  a simple solar panel-based receiver. These strategies are performed 
at the transmitter or at the receiver, or at both sides, named \textit{Adjusting transmission}, \textit{Adjusting reception} and \textit{Coordinated adjustment of transmission and reception}, correspondingly. Furthermore, we deal with the fundamental trade-off between harvested energy and  quality-of-service (QoS),  by maximizing the harvested energy, while achieving the required user's QoS. To this end, two optimization problems are formulated  and optimally solved. Computer simulations validate the optimum solutions and reveal that the proposed strategies considerably increase  the harvested energy, compared to SLIPT with fixed policies.
\end{abstract}

\section{Introduction}

The era of Internet-of-Things (IoT) opens up the opportunity for a number of promising applications in smart buildings,
health monitoring, and predictive maintenance. In the context of wireless access to IoT devices,  radio frequency (RF) technology is the main 
enabler. Furthermore, the exponential growth in the data traffic puts tremendous pressure on the existing 
global telecommunication networks and the expectations from the fifth generation (5G) of wireless networks. However, 
it is remarkable that most of the data consumption/generation, which are related to IoT applications, occurs in 
indoor environments \cite{Volker}. Motivated by this, optical wireless  communicationσ (OWC), 
such as visible light communications (VLC) or  infrared (IR), have been recognized as  promising 
alternative/complimentary technologies to RF, in order to give access to IoT devices in indoor applications \cite{Volker}. 
The data rates reported for indoor VLC/IR networking are much higher than those achieved by WiFi, 
especially when client and server are closely located. Apart from the very high data rates \cite{Hranilovic1}, the advantages of OWC 
technologies include: i) increase of available bandwidth, ii) easy bandwidth reuse, iii) increase of energy efficiency  and 
considerable energy savings, iv) no RF contamination, and v) free from RF interference. Moreover, nowadays, light  emitting diodes (LEDs) and photodetectors (PDs) tend to 
be considerably cheaper than their RF counterparts, while the cost-efficiency is further improved due to the potential to use the 
existing lighting infrastructure \cite{Kaverhard, Arnon1, Arnon2}.

Due to the strong dependence of the IoT on wireless access, their applications are 
constrained by the finite battery capacity of the involved devices \cite{SUde}. Therefore,  energy harvesting (EH), which refers to harnessing energy from the environment or other sources and converting to 
electrical energy, is a critical part of the operation and maintain of the IoT devices. Energy harvesting is regarded as a disruptive technological 
paradigm to prolong the lifetime of energy-constrained wireless networks, which apart from offering a promising solution for 
energy-sustainability of wireless nodes, it also reduces the operating expenses (OPEX) \cite{SUde}. However, the main disadvantage of traditional EH methods is that they rely on natural resources, such as solar and wind, which are uncontrollable. For this reason, harvesting energy from radio frequency signals, which also transfer information, seems to be an interesting alternative. In order to enable simultaneous wireless information and power transfer and increase efficiency of the utilized resources, two strategies have been proposed named \textit{power-splitting}, which is based on the division of the signal’s power into two streams, and \textit{time-splitting}, according to which, during a portion of time, the received signal is used solely for energy harvesting, instead of decoding \cite{book}.

Although RF based wireless power transfer is a well investigated topic in the last five years, optical wireless power transfer (OWPT)  is a new topic and only a few works have been reported so far in the open literature.  In the pioneering work of Fakidis et. al. \cite{Fakidis}, the visible and  infra-red parts of the electromagnetic (EM) spectrum was  used for OWPT, through laser or LEDs at the transmitter and solar cells at the receiver side. Also, in \cite{carvalho} and \cite{Nasiri} energy harvesting was performed by using the existing lighting fixtures for indoor IoT applications. Regarding the simultaneous optical wireless information and power transfer, in \cite{Li} the sum rate maximization
problem has been optimized in a downlink VLC system with simultaneous wireless information and power transfer.  However, in this paper the utilized energy harvesting model 
does not correspond to that of the solar panel, where only the direct current DC component of the modulated light can be used for energy harvesting, in contrast to the alternating current AC component, which carries the information. The separation of the DC and AC  components was efficiently achieved by the self-powered solar panel receiver proposed in \cite{Haas1, Haas2}, where it was proved that the use of the solar panel  for communication purposes does not limit its energy harvesting capabilities. Thus, the utilization of the power-splitting in the useful recent work \cite{sandalidis}, where the received photocurrent is splitted  in two parts with each of them to include  both a 
DC and a AC part, reduces the EH efficiency. Moreover, in \cite{sandalidis} an oversimplified energy harvesting model was used, assuming that the harvested energy is linearly proportional to the received optical power, while an optimization of the splitting technique was not presented. Furthermore, in the significant research works \cite{Alouini, Alouini2}, a dual-hop hybrid VLC/RF communication system is considered, in order to extend the coverage. In these  papers, besides detecting the information over the VLC link, the relay is also able to harvest energy from the first-hop VLC link, by extracting the DC component of the received optical signal. This energy can be used to re-transmit the data to a mobile terminal over the second-hop RF link. Also,  in \cite{Alouini} the proposed hybrid system was optimized, in terms of data rate maximization, while in \cite{Alouini2} the packet loss probability was evaluated.

In this paper, we present for first time a framework for simultaneous optical wireless information and power transfer, called from now on as  \textit{Simultaneous Lightwave Information and Power Transfer (SLIPT)}, which 
can be efficiently used for indoor IoT applications through VLC or IR systems. More specifically, we propose novel 
and fundamental strategies in order to increase the feasibility and efficiency of SLIPT, when a solar panel-based receiver is used. These strategies are performed at the transmitter or at the receiver, or at both sides, named \textit{Adjusting transmission}, \textit{Adjusting reception}, and \textit{Coordinated adjustment of transmission and reception}. Regarding adjusting transmission 
two policies are proposed:  i) \textit{Time-splitting (TS)}, according to which the time frame is separated in two 
distinct phases, where in each of them the main focus is either on communication  or energy transfer and,
ii) \textit{Time-splitting with DC bias optimization}, which is a generalization of TS. In contrast to RF-based wireless 
powered networks, where the TS strategy and adjustment of the related parameters takes place at the receiver's side, 
TS in SLIPT refers to the adaptation of specific parameters of the transmitted signal. Regarding adjusting reception, the 
\textit{Field-of-view (FoV) adjustment} policy is proposed, while according to the coordinated adjustment of transmission and reception strategy, we propose the simultaneous optimization of the former policies at both transmitter and receiver, in order to maximize the harvested energy, while achieving the required Quality-of-Service (QoS) (e.g.  data rate and 
signal-to-noise plus interference ratio (SINR)). Finally, the resulting two optimization problems are formulated and optimally solved.

\section{System and Channel model}\label{S:Intro}
\begin{figure}
\centering
\includegraphics[width=0.85\columnwidth]{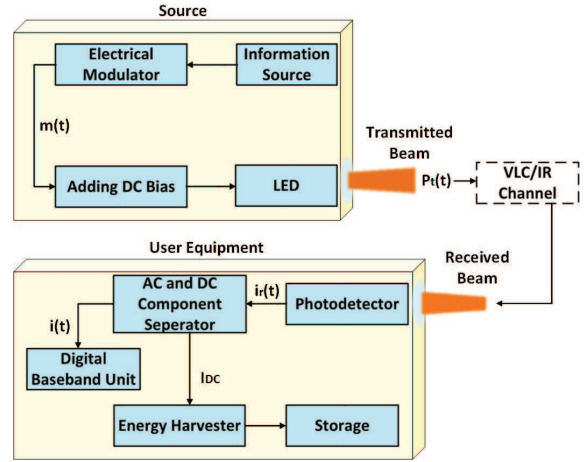}
\vspace{-0.05 in}
\caption{SLIPT transceiver design}
\label{Fig1}
\end{figure}

\begin{figure}
\centering
\includegraphics[width=0.85\columnwidth]{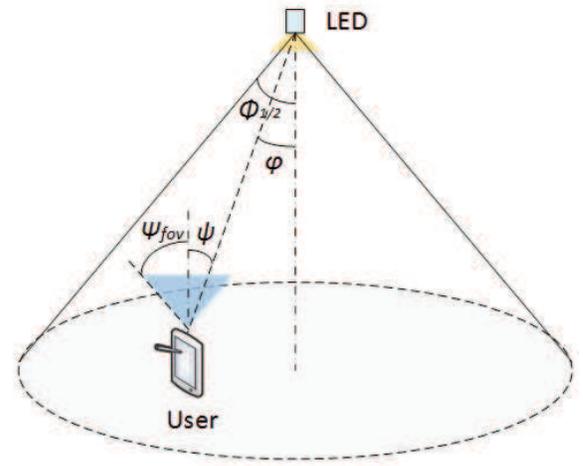}
\vspace{-0.05 in}
\caption{VLC/IR downlink communication}
\label{Fig2}
\end{figure}

We consider the downlink transmission of an OWC system, consisting of one LED and a single user. We also assume that the user is equipped with the functionality of energy harvesting. The VLC/IR transmitter/receiver design is shown in Fig. \ref{Fig1}, while the VLC/IR downlink communication is depicted in Fig. \ref{Fig2}.

\subsection{Optical Wireless Transmission}
Let $m(t)$ denote the modulated electrical signal that corresponds to the bit stream from the information source. A DC 
bias $B$ is added to $m(t)$ to ensure that the resulting signal is non-negative,  before being used to 
modulate the optical intensity of the LED and regulate the LED in the proper operation mode.
The transmitted optical signal from the LED is \cite{Alouini}
\begin{equation}
P_{t}(t)=P_\mathrm{LED}[B+m(t)],
\end{equation}
where $P_\mathrm{LED}$ is the LED power.
The electrical signal varies around the DC bias $B\in[I_L,I_H]$ with peak amplitude  $A$, where $I_L$ is the minimum and $I_H$ is the maximum input bias currents, correspondigly.  In order to avoid clipping distortion by the nonlinearity of the LED, by restraining the input electrical signal to the LED within the linear region of the LED operation, the following limitation is induced
\begin{equation}
A\leq\min(B-I_L,I_H-B),
\label{constraint}
\end{equation}
where $\min(z,y)$ denotes the minimum between $z$ and $y$.

\subsection{Channel Model}
The channel power gain is given by \cite{Komine,Hranilovic2, Kahn}
\begin{equation}\label{channelpowergain}
h=\frac{L_r}{d^2}R_0(\varphi)T_s(\psi)g(\psi)\cos(\psi),
\end{equation}
where $L_r$ is the physical area of
the photo-detector, $d$ is the transmission distance from the LED to the illuminated
surface of the photo-detector, $T_s(\psi)$ is the gain of the optical filter and $g(\psi)$ represents the gain of the optical concentrator, given by \cite{Komine,Kahn}
\begin{equation}
g(\psi)=\begin{cases}
\frac{\rho^2}{\sin^2 (\Psi_\mathrm{fov})},\,0\leq \psi\leq \Psi_\mathrm{fov},\\
0,\, \psi> \Psi_\mathrm{fov}.
\end{cases}
\end{equation}
with $\rho$ and $\Psi_\mathrm{fov}$ being the refractive index and FOV, respectively.
Also in (\ref{channelpowergain}),  $R_0(\varphi)$ is the Lambertian radiant intensity of the LED, given by
\begin{equation}
R_0(\varphi)=\frac{\xi+1}{2\pi}\cos^\xi \varphi,
\end{equation}
where $\varphi$ is the irradiance angle, $\psi$ is the incidence angle, and
\begin{equation}
\xi=-\frac{1}{\log_2\cos(\Phi_{1/2})},
\end{equation}
with $\Phi_{1/2}$ being the semi-angle at half luminance.

\subsection{Received Electrical SINR}
The electrical current $i_r(t)$ at the output of the PD can be written as
\begin{equation}
i_r=\eta (h P_{t}(t)+P_{o})+n(t)=I_{\mathrm{DC}}(t)+i(t)+n(t),
\end{equation}
where $\eta$ is the photo-detector responsivity in $\mathrm{A/W}$, $P_o$ is the received optical signal from other sources, e.g. other neighboring LEDs, $I_{\mathrm{DC}}$ is the DC component, $i(t)$ is the AC component, and $n(t)$ is the additive white Gaussian noise (AWGN), which is created from background shot noise and thermal noise.

The AC component $i(t)$ is composed of two terms, i.e. $i(t)=i_1(t)+i_2(t)$, where
\begin{equation}
i_1(t)=\eta h P_\mathrm{LED}m(t)
\end{equation}
is due to the dedicated LED, and $i_2(t)$ is due to other interfering sources.
Thus, the received SINR can be written as
\begin{equation}
\gamma=\frac{(\eta h P_\mathrm{LED}A)^2}{P_I+\sigma^2},
\label{SINR}
\end{equation}
where $\sigma^2$ is the noise power and $P_I$ is the electrical power of the received interference.

\subsection{Energy Harvesting Model}
As it has already been mentioned the photocurrent consistes of both the DC and AC signals. In order to perform energy harvesting, the DC component is blocked by a capacitor and passes through the energy harvesting branch \cite{Haas1}. The harvested energy is given by \cite{li2011solar}
\begin{equation}
E=fI_\mathrm{DC}V_\mathrm{oc},
\label{energy harvesting}
\end{equation}
with $f$ being the fill factor \cite{li2011solar} and $I_\mathrm{DC}=I_1+I_2$
being the DC component of the output current, where
\begin{equation}
I_1=\eta h_nP_\mathrm{LED}B
\end{equation}
is due to the dedicated LED, while $I_2$ is due to different light sources, e.g. neighboring LEDs. Also, $V_{\mathrm{oc}}$ is 
\begin{equation}
V_{\mathrm{oc}}=V_t\ln(1+\frac{I_{\mathrm{DC}}}{I_0}),
\end{equation}
where $V_t$ is the thermal voltage and $I_0$ is the dark saturation current of the
PD. Moreover, $f$ is the fill factor, defined as the ratio of the maximum power from the solar cell to the product of the open-circuit voltage $V_\mathrm{oc}$ and $I_\mathrm{sc}$,

\section{SLIPT Strategies}
In this section we propose fundamental SLIPT strategies for use in VLC/IR communication systems. These strategies are  performed either at the transmitter  or at the receiver, or at both sides: \textit{Adjusting transmission}, \textit{Adjusting reception}, and \textit{Coordinated adjustment of transmission and reception}.

\subsection{Adjusting Transmission}
Next, we introduced two policies for the adjusting transmission strategy, named \textit{Time-splitting} and  \textit{Time-splitting with DC Bias Optimization}.
\subsubsection{Time-splitting}
\label{strategy 1}
According to the Time-splitting policy the received optical signal is  used for a portion of time solely for  energy harvesting, instead of decoding. During this period of time the LED  transmits by using the maximum DC bias, in order to maximize the harvested energy by the receiver. Thus, assuming time frames of unitary duration, there are the following two distinct phases during a time frame:

\underline{Phase $1$}: The AC component of the received signal is used for information decoding and the DC component for energy harvesting. Let  $A_1$ and $B_1\in[I_L,I_H]$ denote the peak amplitude of $m(t)$ and DC bias, respectively. During Phase 1, the aim is to maximize the received SINR.  Since SINR is an increasing function with respect to $A_1$, then $A_1$ takes its maximum value, which, considering \eqref{constraint} is given by $A_1=\frac{I_H-I_L}{2}$ and similarly, $B_1=\frac{I_H+I_L}{2}$. The duration of this phase is denoted by $0\leq T\leq 1$, which can be optimized according to the QoS requirements. For a specific value of $T$, the amount of harvested energy is given by
\begin{equation}
\begin{split}
E_{TS}^{[1]}&=fT(\eta hP_\mathrm{LED}\frac{I_H+I_L}{2}+I_2)V_t\\
&\times \ln(1+\frac{\eta hP_\mathrm{LED}\frac{I_H+I_L}{2}+I_2}{I_0}).\\
\end{split}
\end{equation}

\underline{Phase $2$}: In the time period $1-T$, the aim is to maximize the harvested energy, which is an increasing function with respect to $B$. Thus, during Phase 2 the transmitter eliminates the AC part and maximizes the DC bias, i.e., $A=0$ and $B=I_H$, where  $A_2$ and $B_2\in[I_L,I_H]$ denote the values of $A$ and $B$, respectively. 
Thus, the amount of harvested energy during this phase, is given by
\begin{equation}
\begin{split}
E_{TS}^{[2]}=&f(1-T)(\eta hP_\mathrm{LED}I_H+I_2)V_t\times\\
&\ln(1+\frac{\eta hP_\mathrm{LED}I_H+I_2}{I_0}).
\end{split}
\end{equation}

Considering both phases, the total harvested energy is given by
\begin{equation}
E_{TS}=E_{TS}^{[1]}+E_{TS}^{[2]}.
\label{harvested energy TS}
\end{equation}

\subsubsection{Time-Splitting with DC Bias Optimization}
\label{strategy 2}
This policy is a generalization of Time-splitting. During Phase $1$, the DC bias is optimized in order to increase the harvested energy, while it simultaneously enables information transfer, i.e., $A_1> 0$. In this case, the total harvested energy is given by
\begin{equation}
\begin{split}
E_\mathrm{TSBO}=&fT(\eta hP_\mathrm{LED}B_1+I_2)V_t\times\\
&\ln(1+\frac{\eta hP_\mathrm{LED}B_1+I_2}{I_0})+E_{TS}^{[2]},\\
\end{split}
\label{harvested energy TSBO}
\end{equation}
where $B_1$ is the DC bias during Phase $1$.

\subsection{Adjusting Reception}
\label{strategy 3}
We propose the \textit{Adjustment of the field of view (FOV)} policy for the adjusting reception strategy, in order  to balance the trade-off between  harvested energy and SINR. Controlling FOV is particularly important especially when, except for the used VLC/IR LED, there are extra light sources in the serving area \cite{Ghamdi}, e.g. neighboring LEDs that serve other users. For the practical and efficient implementation of this policy, electrically controllable liquid crystal (LC) lenses is a promising technology \cite{Tsou}.

When the aim is to maximize the SINR, the FOV is tuned up to receive the beam of the dedicated LED only (if possible), in order to reduce the beam overlapping. This is achieved by tuning the FOV to the narrowest setting, that allows reception only from that LED. On the other hand, when the aim is to achieve a balance between SINR and harvested energy, a wider FOV setting could be selected. 

For the sake of practicality, we assume that the VLC/IR receiver has discrete FOV settings, i.e. $\Psi_\mathrm{fov}\in\{\Psi_\mathrm{fov}^{[1]},..,\Psi_\mathrm{fov}^{[M]}\}$. Also, note that except for $h$, both $P_I$ and $I_2$ are also discrete functions of $\Psi_\mathrm{fov}$, i.e., $P_I=P_I(\Psi_\mathrm{fov})$ and $I_2=I_2(\Psi_\mathrm{fov})$.

\subsection{Coordinated Transmission and Reception Adjustment}
\label{strategy 4}
Considering \eqref{SINR}, \eqref{harvested energy TS}, and \eqref{harvested energy TSBO}, it is revealed that both   SINR and  harvested energy -apart from $A_1$, $B_1$ and $T$-  also depend on the selection of $\Psi_\mathrm{fov}$, despite the utilized adjusting transmission technique. This dependence motivates the coordinated transmission and reception adjustment, i.e. the coordination between the strategy \ref{strategy 1} or \ref{strategy 2} and \ref{strategy 3}, which results in the following two policies, i.e.
\begin{itemize}
\item Policy 1: Time-splitting with tunable FOV (\ref{strategy 1} and \ref{strategy 3})
\item Policy 2: Time-splitting with DC bias optimization and tunable FOV (\ref{strategy 2} and \ref{strategy 3})
\end{itemize}
Note that in both policies, during Phase 2, where the aim is to maximize the harvested energy, the FOV setting that maximizes $E_{TS}^{[2]}$ should be used. This is not necessarily the widest setting, because although it increases the received beams (if there are neighboring LEDs), it reduces $g(\psi)$. On the other hand, the preferable FOV setting during phase 1, denoted by $\Psi_\mathrm{fov,1}$, cannot be straightforwardly determined, since it also depends  on the required QoS.

\section{SLIPT Optimization}

SLIPT induces an interesting trade-off between harvested energy and  QoS. In this section, we aim to balance this trade-off by maximizing the harvested energy, while achieving the required user QoS. In the present work, we focus on the coordinated adjustment of transmission and reception strategy, which can be considered as a generalization of the other SLIPT strategies. The following optimization problems can be formulated, based on the two techniques presented in subsection \ref{strategy 4}.

Regarding the QoS, two different criteria are taken into account, namely SINR and information rate. Note that these two criteria are not equivalent to each other, when either of the two techniques is used, due to the time-splitting.  More specifically,  since only Phase $1$ is used for information transmission (the duration of which is $T$), the lower bound of the capacity is given by \cite{Wangrate}
\begin{equation}
R=T\log_2\left(1+\frac{e}{2\pi}\gamma\right).
\label{rate}
\end{equation}

\subsection{Time-Splitting with Tunable FOV}
The corresponding optimization problem can be expressed as

\begin{equation}
\begin{array}{ll}
\underset{T, \Psi_\mathrm{fov,1}}{\text{\textbf{max}}}&  E_h^\mathrm{TS} \\
\text{\textbf{s.t.}}&\mathrm{C}_1: R\geq R_{\mathrm{th}},\\
& \mathrm{C}_2: \gamma\geq \gamma_{\mathrm{th}},\\
&\mathrm{C}_3:0 \leq T\leq 1 ,\\
&\mathrm{C}_4:\Psi_\mathrm{fov,1}\in\{\Psi_\mathrm{fov}^{[1]},..,\Psi_\mathrm{fov}^{[M]}\},\\
\end{array}
\label{opt TS}
\end{equation}
where $R_{\mathrm{th}}$ and $\gamma_{\mathrm{th}}$ denote the information rate SINR and threshold, respectively. 
\begin{theorem}
The optimal value of $T$ in (\ref{opt TS}) is given by
\begin{equation}
T^*=\frac{R_{\mathrm{th}}}{\log_2\left(1+\frac{e\left(\eta h P_\mathrm{LED}(I_H-I_L)\right)^2}{8\pi(P_I(\Psi^*_\mathrm{fov,1})+\sigma^2)}\right)},
\label{optimal T TS}
\end{equation}
where $(\cdot)^*$ denotes optimality.
\end{theorem}

\begin{IEEEproof}
The optimization problem \eqref{opt TS} is a combinatorial one. In order to find the optimal solution, all possible values of $\Psi_\mathrm{fov,1}$  have to be checked before selecting the value that maximizes the harvested energy, $E_h^\mathrm{TS}$, while satisfying the constraints $\mathrm{C}_1$, $\mathrm{C}_2$, and $\mathrm{C}_3$. For a specific specific value of $\Psi_\mathrm{fov,1}$, if
\begin{equation}
\frac{(\eta h P_\mathrm{LED}\frac{I_H-I_L}{2})^2}{P_I(\Psi_\mathrm{fov,1})+\sigma^2}<\gamma_{\mathrm{th}},
\end{equation}
then the optimization problem is infeasible, since $\mathrm{C}_2$ is not satisfied.
Also, due to constraint $\mathrm{C}_1$, the following limitation is induced for $T$, 
\begin{equation}
T \geq\frac{R_{\mathrm{th}}}{\log_2\left(1+\frac{e(\eta h P_\mathrm{LED}\frac{I_H-I_L}{2})^2}{2\pi(P_I(\Psi_\mathrm{fov,1})+\sigma^2)}\right)}.
\end{equation}
Moreover, the harvested energy is decreasing with respect to $T$. Thus, the optimal value of $T$ is given by
\eqref{optimal T TS} and the proof is completed.
\end{IEEEproof}
Note that if $T^*>1$, the optimization problem in \eqref{opt TS} is infeasible, due to $\mathrm{C}_3$.

\subsection{Time-Splitting with DC Bias Optimization and tunable FOV}

The corresponding optimization problem can be formulated as

\begin{equation}
\begin{array}{ll}
\underset{B_1, A_1, T, \Psi_\mathrm{fov,1}}{\text{\textbf{max}}}&  E_h^\mathrm{TSBO} \\
\text{\textbf{s.t.}}&\mathrm{C}_1: R\geq R_{\mathrm{th}},\mathrm{C}_2: \gamma\geq \gamma_{\mathrm{th}},\\
&\mathrm{C}_3:A_1\leq \min(B_1-I_L,I_h-B_1),\\
&\mathrm{C}_4:0 \leq T\leq 1 ,\\
&\mathrm{C}_5:A_1\geq 0,\mathrm{C}_6:I_L\leq B_1\leq I_H,\\
&\mathrm{C}_7:\Psi_\mathrm{fov,1}\in\{\Psi_\mathrm{fov}^{[1]},..,\Psi_\mathrm{fov}^{[M]}\}.\\
\end{array}
\label{opt TS DC}
\end{equation}
\begin{proposition}
The optimal value of $B$ in (\ref{opt TS DC}) belongs in the range $\left[\frac{I_H+I_L}{2},I_H\right]$.
\label{Proposition1}
\end{proposition}

\begin{IEEEproof}
The constraint $\mathrm{C}_3$ can be rewritten as
\begin{equation}
\mathrm{C}_{3\mathrm{a}}:A_1\leq B_1-I_L, \mathrm{C}_{3\mathrm{b}}:A_1\leq I_H-B_1.
\end{equation}
For a specific value of $B_1$, only one of the constraints $\mathrm{C}_{3\mathrm{a}}$ and $\mathrm{C}_{3\mathrm{b}}$ is activated. Now, let assume that the optimal solution is  $B_1^*<\frac{I_H+I_L}{2}$, for which all the constraints are satisfied. In this case, $\mathrm{C}_{3\mathrm{a}}$ is activated.  However, by setting $B_1=\frac{I_H+I_L}{2}$ the objective function is increased, while the constraints are still satisfied. Thus, we can infer that that $B_1^*$ is not optimal. Consequently, Proposition \ref{Proposition1} has been proved by contradiction.
\end{IEEEproof}

The optimal value $\Psi_\mathrm{fov,1}$ is calculated similarly to the solution  of \eqref{opt TS}. Regarding the rest optimization variables of \eqref{opt TS DC} they are optimized according to the following theorem:
\begin{theorem}
\label{Theorem2}
For a specific value of $\Psi_\mathrm{fov,1}$, the optimal value of $T$ is given by
\begin{equation}
T^*=\underset{K_1\leq T\leq K_2}{\argmax} \tilde{E}_h^\mathrm{TSBO}
\label{equivalent}
\end{equation}
with $\tilde{E}_h^\mathrm{TSBO}$ being solely a function of $T$ and given by \eqref{harvested energy TSBO}, by replacing $A_1$ and $B_1$ by
\begin{equation}
A_1=\frac{1}{nhP_\mathrm{LED}}\sqrt{\frac{2\pi(P_I(\Psi_\mathrm{fov,1})+\sigma^2)(2^{\frac{R_\mathrm{th}}{T}}-1)}{e}},
\label{A1}
\end{equation}
and
\begin{equation}
B_1=I_H-A_1,
\label{B1}
\end{equation}
respectively.
Also,
\begin{equation}
K_1=\frac{R_\mathrm{th}}{\log_2\left(1+\frac{e\left(\eta h_{1}P_\mathrm{LED}(I_H-I_L)\right)^2}{8\pi\left(P_I(\Psi^*_\mathrm{fov,1})+\sigma^2\right)}\right)}
\end{equation}
\begin{equation}
K_2=\min\left(\frac{R_\mathrm{th}}{\log_2\left(1+\frac{e\gamma_\mathrm{th}}{2\pi}\right)},1\right).
\end{equation}
Finally, the optimal values of $A_1$ and $B_1$ are given by \eqref{A1} and \eqref{B1}, by replacing $\Psi_\mathrm{fov,1}$ and $T$ by $\Psi^*_\mathrm{fov,1}$ and $T^*$, respectively.
\end{theorem}

\begin{IEEEproof}
Considering Proposition \ref{Proposition1} and for a specific value of $\Psi_\mathrm{fov,1}$ the optimization problem in \eqref{opt TS DC} can be re-formulated as
\begin{equation}
\begin{array}{ll}
\underset{B_1, A_1, T}{\text{\textbf{max}}}&  E_h^\mathrm{TSBO} \\
\text{\textbf{s.t.}}&\mathrm{C}_1: R\geq R_{\mathrm{th}},\mathrm{C}_2: \gamma\geq \gamma_{\mathrm{th}},\\
&\mathrm{C}_3:A_1+B_1\leq I_H ,\mathrm{C}_4:0 \leq T\leq 1,\\
&\mathrm{C}_5:A_1\geq 0,\mathrm{C}_6:B_1\geq \frac{I_H+I_L}{2}.\\
\end{array}
\label{reformulated}
\end{equation}

The optimization problem in \eqref{reformulated} still cannot be easily solved in its current form, since the objective function as well as the constraints $\mathrm{C}_1$ and $\mathrm{C}_2$ are not concave. However, it can be solved with low complexity by using the following reformulation.

First, the inequalities in $\mathrm{C}_1$ and  $\mathrm{C}_3$ are replaced by equalities. Then, $A_1$ and $B_1$ are given by \eqref{A1} and \eqref{B1}, respectively. By substituting $T_1$ and $B_1$ by \eqref{A1} and \eqref{B1}, $\mathrm{C}_1$,  $\mathrm{C}_3$, and $\mathrm{C}_3$ of \eqref{reformulated} vanish, and the optimization problem is rewritten as

\begin{equation}
\begin{array}{ll}
\underset{B_1, A, T_1\forall n}{\text{\textbf{max}}}&  \tilde{E}_h^\mathrm{TSBO} \\
\text{\textbf{s.t.}}& \mathrm{C}_2: T\leq \frac{R_\mathrm{th}}{\log_2\left(1+\frac{e\gamma_\mathrm{th}}{2\pi}\right)},\\
&\mathrm{C}_4:0\leq T\leq 1 ,\\
&\mathrm{C}_6:T\geq \frac{R_\mathrm{th}}{\log_2\left(1+\frac{e\left(\eta h_{1}P_\mathrm{LED}(I_H-I_L)\right)^2}{8\pi\left(P_I+\sigma^2\right)}\right)},\\
\end{array}
\label{opt NOMA general}
\end{equation}
which is equivalent to \eqref{equivalent}, and, thus, the proof is completed.

\end{IEEEproof}

\vspace{-0.1 in}
\section{Simulations and discussion}

We assume the downlink VLC/IR system of Fig. \ref{Fig2}, where  the user is located in a distance $d=1.5$ m from the LED, $\psi=0$, and the transmitter plane is parallel to the receiver one, i.e., $\varphi=\psi$. In the same room there are $N$ other LEDs, which simultaneously use the same frequency band. The distance between each of them and from the dedicated LED is $D=1.5$ m. We also assume $f=0.75$, $P_{LED}=20$ W/A, $\Phi_{1/2}=60$ deg, $\sigma^2=10^{-15}$ A$^2$, $L_r=0.04$ m$^2$, $\eta=0.4$ A/W, $I_0=10^{-9}$ A, $I_L=0$ A, $I_H=12$ mA \cite{Alouini}, $T_s=1$, $\rho=1.5$, $\gamma_{th}=10$ dB, and  two settings for the FOV, i.e., $\Psi_\mathrm{fov}\in\{30,50\}$ deg, are considered. 

Regarding the neighboring LEDs, we assume that the DC bias and the peak amplitude are given by $A_n'=B_n'=6$ mA, $\forall n\in\{1,...,N\}$, while the rest parameters are equal to those of the dedicated LED. Furthermore, the channel between them and the user's receiver, denoted by $h_n$ is modeled according to \eqref{channelpowergain}, using the corresponding parameters. Thus, when the widest FOV setting is selected, $P_I$ and $I_2$ are given by
\begin{equation}
P_I=\sum_{n=1}^N(\eta h_nP_{LED}A_n')^2
\end{equation}
and
\begin{equation}
I_2=\sum_{n=1}^N\eta h_nP_{LED}B_n',
\end{equation}
otherwise their values are zero.

\begin{figure}[t!]
\centering
\includegraphics[width=0.85\columnwidth]{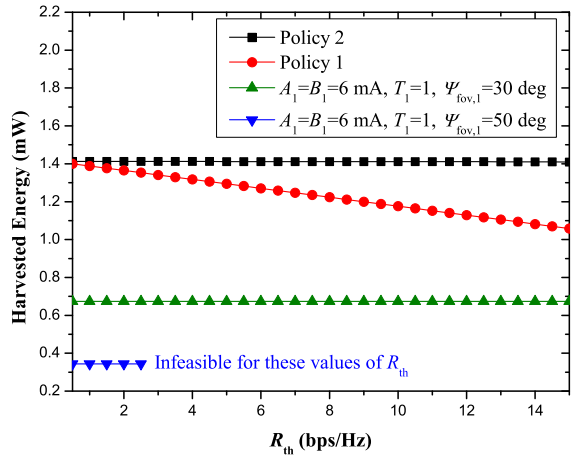}
\caption{Harvested energy vs $R_{th}$ for $N=1$.}
\label{Fig3}
\end{figure}

\begin{figure}[t!]
\centering
\includegraphics[width=0.85\columnwidth]{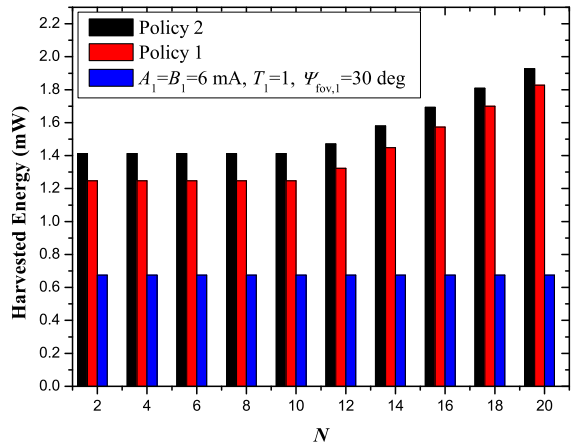}
\caption{Harvested energy vs $N$ for $R_{th}=7$ bps/Hz.}
\label{Fig4}
\end{figure}

The performance of both optimized policies of Section \ref{strategy 4} are compared for $N=1$, while they are also presented against the case of fixed $A_1$, $B_1$, $T_1$, and $\Psi_{\mathrm{fov},1}$, which is considered as the baseline policy. More specifically, in Fig. \ref{Fig3} the harvested energy is plotted against the rate threshold. As it is observed, both policies significantly outperform the baseline for both values of $\Psi_{\mathrm{fov},1}$. Regarding the baseline, the value $\Psi_{\mathrm{fov},1}=50$ deg reduces the harvested energy compared to $\Psi_{\mathrm{fov},1}=30$ deg, because $g(\psi)$ decreases and thus,  cancels the benefit of receiving the beam of the neighboring LED. Also, the baseline policy with $\Psi_{\mathrm{fov},1}=50$  deg is infeasible  for medium and high values of $R_{th}$, because the rate threshold cannot be reached, due to the received interference. Interestingly, Policy 2 outperforms Policy 1, especially for the high region of $R_{th}$, which is due to the extra degrees of freedom. Similar conclusions can be obtained by Fig. \ref{Fig4}, where the harvested energy is plotted against the number of neighboring LEDS. For this specific setup, the baseline with $\Psi_{\mathrm{fov},1}=50 $ deg is not feasible, and, thus, it is omitted. We notice here that for a small number of neighboring LEDs, the harvested energy remains constant, since the receiver prefers the smallest FOV setting. However, as the number of neighboring LEDs increases, the receiver prefers the widest FOV setting and the harvested energy increases with the increase of LEDs.


\begin{thebibliography}{10}
\providecommand{\url}[1]{#1}
\csname url@samestyle\endcsname
\providecommand{\newblock}{\relax}
\providecommand{\bibinfo}[2]{#2}
\providecommand{\BIBentrySTDinterwordspacing}{\spaceskip=0pt\relax}
\providecommand{\BIBentryALTinterwordstretchfactor}{4}
\providecommand{\BIBentryALTinterwordspacing}{\spaceskip=\fontdimen2\font plus
\BIBentryALTinterwordstretchfactor\fontdimen3\font minus
  \fontdimen4\font\relax}
\providecommand{\BIBforeignlanguage}[2]{{%
\expandafter\ifx\csname l@#1\endcsname\relax
\typeout{** WARNING: IEEEtran.bst: No hyphenation pattern has been}%
\typeout{** loaded for the language `#1'. Using the pattern for}%
\typeout{** the default language instead.}%
\else
\language=\csname l@#1\endcsname
\fi
#2}}
\providecommand{\BIBdecl}{\relax}
\BIBdecl

\bibitem{Volker}
M.~Ayyash, H.~Elgala, A.~Khreishah, V.~Jungnickel, T.~Little, S.~Shao,
  M.~Rahaim, D.~Schulz, J.~Hilt, and R.~Freund, ``Coexistence of {W}i{F}i and
  {L}i{F}i toward 5{G}: {C}oncepts, {O}pportunities, and {C}hallenges,''
  \emph{{IEEE} {C}ommun. {M}ag.}, vol.~54, no.~2, pp. 64--71, Feb. 2016.

\bibitem{Hranilovic1}
M.~S.~A. Mossaad, S.~Hranilovic, and L.~Lampe, ``Visible {L}ight
  {C}ommunications {U}sing {OFDM} and {M}ultiple {LED}s,'' \emph{{IEEE}
  {T}rans. {C}ommun.}, vol.~63, no.~11, pp. 4304--4313, Nov. 2015.

\bibitem{Kaverhard}
M.~Kavehrad, ``Sustainable {E}nergy-{E}fficient {W}ireless {A}pplications
  {U}sing {L}ight,'' \emph{{IEEE} {C}ommun. {M}ag.}, vol.~48, no.~12, pp.
  66--73, Dec. 2010.

\bibitem{Arnon1}
S.~Arnon, Ed., \emph{Visible {L}ight Communication}.\hskip 1em plus 0.5em minus
  0.4em\relax Cambridge University Press, 2015.

\bibitem{Arnon2}
D.~Bykhovsky and S.~Arnon, ``Multiple {A}ccess {R}esource {A}llocation in
  {V}isible {L}ight {C}ommunication {S}ystems,'' \emph{J. {L}ightw.
  {T}echnol.}, vol.~32, no.~8, pp. 1594--1600, Apr. 2014.

\bibitem{SUde}
S.~Sudevalayam and P.~Kulkarni, ``Energy {H}arvesting {S}ensor {N}odes:
  {S}urvey and {I}mplications,'' \emph{{IEEE} {C}ommun. {S}urveys {T}uts.},
  vol.~13, no.~3, pp. 443--461, Jul. 2010.

\bibitem{book}
S.~Nikoletseas, Y.~Yang, and A.~Georgiadis, Eds., \emph{Wireless {P}ower
  {T}ransfer {A}lgorithms, {T}echnologies and {A}pplications in {A}d {H}oc
  {C}ommunication {N}etworks}.\hskip 1em plus 0.5em minus 0.4em\relax Springer,
  2016.

\bibitem{Fakidis}
J.~Fakidis, S.~Videv, S.~Kucera, H.~Claussen, and H.~Haas, ``Indoor {O}ptical
  {W}ireless {P}ower {T}ransfer to {S}mall {C}ells at {N}ighttime,'' \emph{J.
  Lightw. Technol.}, vol.~34, no.~13, pp. 3236--3258, Jul. 2016.

\bibitem{carvalho}
C.~Carvalho and N.~Paulino, ``On the {F}easibility of {I}ndoor {L}ight {E}nergy
  {H}arvesting for {W}ireless {S}ensor {N}etworks,'' \emph{Procedia
  Technology}, vol.~17, pp. 343--350, 2014.

\bibitem{Nasiri}
A.~Nasiri, S.~A. Zabalawi, and G.~Mandic, ``Indoor {P}ower {H}arvesting using
  {P}hotovoltaic {C}ells for {L}ow-{P}ower {A}pplications,'' \emph{{IEEE}
  {T}rans. {I}nd. {E}lectron.}, vol.~56, no.~11, pp. 4502--4509, Nov. 2009.

\bibitem{Li}
Y.~Li, N.~Huang, J.~Wang, Z.~Yang, and W.~Xu, ``Sum {R}ate {M}aximization for
  {VLC} {S}ystems with {S}imultaneous {W}ireless {I}nformation and {P}ower
  {T}ransfer,'' \emph{{IEEE} {P}hoton. {T}echnol. {L}ett.}, vol.~29, no.~6, pp.
  531--534, Mar. 2017.

\bibitem{Haas1}
Z.~Wang, D.~Tsonev, S.~Videv, and H.~Haas, ``Towards {S}elf-{P}owered {S}olar
  {P}anel {R}eceiver for {O}ptical {W}ireless {C}ommunication,'' in \emph{Proc.
  IEEE International Conference on Communications (ICC)}, Jun. 2014, pp.
  3348--3353.

\bibitem{Haas2}
------, ``On the {D}esign of a {S}olar-{P}anel {R}eceiver for {O}ptical
  {W}ireless {C}ommunications with {S}imultaneous {E}nergy {H}arvesting,''
  \emph{{IEEE} {J}. Sel. Areas Commun.}, vol.~33, no.~8, pp. 1612--1623, Aug.
  2015.

\bibitem{sandalidis}
H.~Sandalidis, A.~Vavoulas, T.~Tsiftsis, and N.~Vaiopoulos, ``Illumination,
  {D}ata {T}ransmission and {E}nergy {H}arvesting: {T}he {T}hreefold
  {A}dvantage of {VLC},'' 2017.

\bibitem{Alouini}
T.~Rakia, H.~C. Yang, F.~Gebali, and M.~S. Alouini, ``Optimal {D}esign of
  {D}ual-{H}op {VLC/RF} {C}ommunication {S}ystem with {E}nergy {H}arvesting,''
  \emph{{IEEE} {C}omm. {L}ett.}, vol.~20, no.~10, pp. 1979--1982, Oct. 2016.

\bibitem{Alouini2}
------, ``Dual-{H}op {VLC}/{RF} {T}ransmission {S}ystem with {E}nergy
  {H}arvesting {R}elay under {D}elay {C}onstraint,'' in \emph{Proc. {IEEE}
  {G}lobecom {W}orkshops}, Dec. 2016, pp. 1--6.

\bibitem{Komine}
T.~Komine and M.~Nakagawa, ``Fundamental {A}nalysis for {V}isible-{L}ight
  {C}ommunication {S}ystem {U}sing {LED} {L}ights,'' \emph{{IEEE} {T}rans.
  {C}onsum. {E}lectron.}, vol.~50, no.~1, pp. 100--107, Feb. 2004.

\bibitem{Hranilovic2}
H.~Ma, L.~Lampe, and S.~Hranilovic, ``Coordinated {B}roadcasting for
  {M}ultiuser {I}ndoor {V}isible {L}ight {C}ommunication {S}ystems,''
  \emph{{IEEE} {T}rans. {C}ommun.}, vol.~63, no.~9, pp. 3313--3324, Sep. 2015.

\bibitem{Kahn}
J.~M. Kahn and J.~R. Barry, ``Wireless {I}nfrared {C}ommunications,''
  \emph{Proc. {IEEE}}, vol.~85, no.~2, pp. 265--298, Feb. 1997.

\bibitem{li2011solar}
C.~Li, W.~Jia, Q.~Tao, and M.~Sun, ``Solar {C}ell {P}hone {C}harger
  {P}erformance in {I}ndoor {E}nvironment,'' in \emph{Proc. IEEE 37th Annual
  Northeast Bioengineering Conference (NEBEC)}, 2011, pp. 1--2.

\bibitem{Ghamdi}
A.~Al-Ghamdi and J.~Elmirghani, ``Performance and {F}ield of {V}iew
  {O}ptimization of an {O}ptical {W}ireless {P}yramidal {F}ly-{E}ye {D}iversity
  {R}eceiver,'' \emph{Journal of optical communications}, vol.~23, no.~6, pp.
  215--222, 2002.

\bibitem{Tsou}
Y.~S. Tsou, Y.~H. Lin, and A.~C. Wei, ``Concentrating {P}hotovoltaic {S}ystem
  using a {L}iquid {C}rystal {L}ens,'' \emph{IEEE Photon. Technol. Lett.},
  vol.~24, no.~24, pp. 2239--2242, Dec. 2012.

\bibitem{Wangrate}
J.~B. Wang, Q.~S. Hu, J.~Wang, M.~Chen, and J.~Y. Wang, ``Tight {B}ounds on
  {C}hannel {C}apacity for {D}immable {V}isible {L}ight {C}ommunications,''
  \emph{J. Lightw. Technol.}, vol.~31, no.~23, pp. 3771--3779, Dec. 2013.

\end{thebibliography}
\end{document}